\documentclass[a4paper,11pt]{article}
\usepackage{pos}

\usepackage{amsmath}
\usepackage{bbm}
\usepackage{bm}
\usepackage{color}
\usepackage{slashed}
\usepackage{graphicx}
\usepackage{dsfont}
\usepackage{bbold}
\usepackage{braket}

\title{Center-of-mass recoil effects on the annihilation and formation of dark matter bound pairs}

\author[a]{S.~Biondini}
\author[b,c,d]{N.~Brambilla}
\author*[b]{G.~Qerimi}
\author[b]{A. Vairo}

\affiliation[a]{Department of Physics, University of Basel,\\
Klingelbergstr. 82, CH-4056 Basel, Switzerland}
\affiliation[b]{TUM School of Natural Sciences, Physics Department, Technical University of Munich,\\
James-Franck-Str.  1, 85748 Garching, Germany}
\affiliation[c]{Institute for Advanced Study, Technical University of Munich, \\
Lichtenbergstrasse 2 a, 85748 Garching, Germany}
\affiliation[d]{Munich Data Science Institute, Technical University of Munich, \\
Walther-von-Dyck-Strasse 10, 85748 Garching, Germany}

\emailAdd{simone.biondini@unibas.ch}
\emailAdd{nora.brambilla@tum.de}
\emailAdd{gramos.qerimi@tum.de}
\emailAdd{antonio.vairo@tum.de}

\abstract{
  For a quantitative investigation on the time evolution of heavy thermal dark matter at and after thermal freeze-out, near-threshold processes need to be taken into account which have a large impact on the observed dark matter relic abundance.
  In this conference paper, we study the recoil effect of heavy dark matter pairs in a thermal bath and compute the annihilation cross section and the decay width as well as the bound-state formation cross section of dark matter fermion-antifermion pairs
  in the laboratory frame within the framework of potential non-relativistic effective field theories at finite temperature.
  For the considered hierarchy of energy scales, we highlight the effect of the recoil corrections to the thermal rates.
}

\FullConference{The European Physical Society Conference on High Energy Physics (EPS-HEP2023)\\
 21-25 August 2023\\
Hamburg, Germany\\}

\begin{document}
\maketitle

\section{Introduction}
\label{sec:intro}
From astrophysical observations it is evident that 80\% of the matter in the cosmos consists of dark matter (DM). Despite its nature being yet unknown, there has been a significant effort in cosmology and particle physics in uncovering its identity and fundamental properties over the last decades. Within particle theory, a prominent scenario which may describe the observed DM relic density, $\Omega_{\hbox{\tiny DM}} h^2 = 0.1200 \pm 0.0012$~\cite{Planck:2018nkj}, is the thermal freeze-out effect. Assuming heavy thermal DM particles moving in a thermal bath modeling the early universe, which expands and hence cools down, under a certain critical temperature the chemical equilibrium condition breaks down while kinetic equilbrium is still maintained. In this proceeding paper, we consider thermal fermionic DM that self-interacts through an abelian long-range mediator field within the dark sector. It is a QED-like model, here dubbed QED$_\text{DM}$. In particular, we study the interactions of the DM particles with the thermal bath of dark photons that can generate the departure from thermal equilibrium and moreover affect the evolution of the DM energy density. Introducing the potential non-relativistic effective field theory for DM pairs in the laboratory frame in section~\ref{sec:pNRQED}, we compute the annihilation cross section and decay width including recoil effects in section~\ref{sec:ann}, and the bound-state formation cross section in section~\ref{sec:bsf}. We comment on the impact of the recoil corrections due to the center-of-mass motion in section~\ref{sec:outlook}. The proceeding is based on ref.~\cite{biondini2024effective}.

\section{pNRQED$_\text{DM}$}
\label{sec:pNRQED}
The Lagrangian density of a dark Dirac fermion $X$ with large mass $M$, charged under an abelian gauge group~\cite{Feldman:2006wd,Fayet:2007ua,Goodsell:2009xc,Morrissey:2009ur,Andreas:2011in}, that we consider is
\begin{equation}
\mathcal{L}_{\textrm{QED}_\text{DM}}=\bar{X} (i \slashed {D} -M) X -\frac{1}{4} F_{\mu \nu} F^{\mu \nu} \, ,
\label{lag_mod_0}
\end{equation}
where $D_\mu=\partial_\mu + i g A_\mu$ is the covariant derivative, $A_\mu$ the dark photon field and $F_{\mu \nu} = \partial_\mu A_\nu - \partial_\nu A_\mu$ the dark field strength tensor. The dark fine structure constant is $\alpha \equiv g^2/(4 \pi)$.

Close-to-threshold processes involve DM fermion-antifermion pairs with non-relativistic relative velocities $v_{\text{rel}} \sim \alpha \ll 1$ that weakly interact with thermal dark photons from the thermal medium of temperature $T$. In addition, upon assuming that the heavy DM particles are thermalized, their momenta scale like $P \sim \sqrt{MT}$. We assume the following hierarchy of energy scales fulfilled for most of the times after chemical decoupling~\cite{Biondini:2023zcz}:\footnote{In a typical freeze-out scenario the decoupling from chemical equilibrium happens around $M/T \approx 25$.}
\begin{equation}
M \gg M\alpha \gtrsim \sqrt{MT} \gg M\alpha^2 \gtrsim T \,.
\label{scale_arrang}
\end{equation}
Since the scales are clearly separated in~\eqref{scale_arrang}, exploiting the formalism of effective field theories (EFTs) allows to integrate out modes of the order of each of the scales subsequently, from highest to lowest, resulting in a tower of non-relativistic EFTs (NREFTs). In our particular model, integrating out hard modes of order $M$ leads to NRQED$_\text{DM}$. Hard processes such as annihilations are encoded in the matching coefficients of the four-fermion local operators. At leading order, those operators are of dimension six and describe S-wave annihilations in the center-of-mass frame of the annihilating pair. In the laboratory frame, however, the pair moves with a non-vanishing thermal total momentum $\bm{P}$, which at leading order in the non-relativistic expansion is represented by suppressed dimension-eight local operators. Hence in the laboratory frame we write~\cite{Brambilla:2008zg,Berwein:2018fos}
\begin{equation}
\begin{aligned}
  & \mathcal{L}_{\textrm{NRQED}_{\textrm{DM}}} ~\supset~
       \frac{d_s}{M^2} \psi^\dagger \chi \, \chi^\dagger \psi
       +\frac{d_v}{M^2} \psi^\dagger \, \bm{\sigma} \, \chi \cdot \chi^\dagger \, \bm{\sigma} \, \psi \\
       &  
       + \frac{g_{\textrm{a\,cm}}}{M^4} \nabla^i(\psi^\dagger \sigma^j \chi) \nabla^i(\chi^\dagger \sigma^j \psi)
         + \frac{g_{\textrm{b\,cm}}}{M^4} \bm{\nabla}\cdot(\psi^\dagger \bm{\sigma} \chi) \bm{\nabla}\cdot(\chi^\dagger \bm{\sigma} \psi)
                  + \frac{g_{\textrm{c\,cm}}}{M^4} \bm{\nabla}(\psi^\dagger \chi) \cdot \bm{\nabla}(\chi^\dagger \psi) \, ,
\label{local_op_total_mom}
\end{aligned}
\end{equation}
where $\psi$ is the two-component Pauli spinor that annihilates a dark fermion, $\chi^\dagger$ is the Pauli spinor that annihilates a dark antifermion
and $\sigma^i$ are the Pauli matrices. The specific values of the matching coefficients $d_s$, $d_v$, $g_{\textrm{a\,cm}}$, $g_{\textrm{b\,cm}}$ and $g_{\textrm{c\,cm}}$ at order $\alpha^2$ can be taken from~\cite{Barbieri:1979be,Hagiwara:1980nv,Brambilla:2008zg,Berwein:2018fos}, and include only spin-singlet pairs annihilating into two dark photons. From Poincar\'e invariance of QED one can deduce general relations such as $\displaystyle \, g_{\textrm{c\,cm}} = -d_s/4$, valid to all orders in the coupling.

Next, we integrate out modes associated to the soft scale $M\alpha$, and multipole expand the dark photon fields in the relative coordinate $\bm{r} \equiv \bm{x}_1-\bm{x}_2 \ll \bm{R}=(\bm{x}_1+\bm{x}_2)/2$ of the pair. The corresponding EFT is called potential NREFT, here dubbed pNRQED$_\textrm{DM}$. Near-threshold processes, such as the formation of a bound state by emitting a thermal or ultrasoft dark photon, are due to electric dipole transitions within the DM pairs, while higher multipole transitions are suppressed at smaller temperatures. Moreover, keeping track of the center-of-mass motion of the pair results in an additional interaction term proportional to the magnetic field, called \textit{R\"ontgen term}~\cite{James_D_Cresser_2003, Brambilla_2003}, that originates from the \textit{Lorentz force} $\bm{F} = \bm{v}\times g\bm{B}$. The Röntgen term is velocity-suppressed compared to the electric-dipole interaction term, with $\bm{v}=\bm{P}/(2M)$. The Lagrangian density in the laboratory frame, that accounts for such processes is given by
\begin{equation}
  \mathcal{L}_{\textrm{pNRQED}_{\textrm{DM}}} ~\supset~   \int d^3r \; \phi^\dagger(t,\bm{r},\bm{R})
             \left[\bm{r} \cdot g\bm{E}(t,\bm{R}) + \bm{r} \cdot \left(\frac{\bm{P}}{2M}\times g\bm{B}(t,\bm{R})\right) + \dots \right] \phi (t,\bm{r},\bm{R})  \,,
\label{pNREFT_1}
\end{equation}
where $E^i=F^{i0}$ is the dark electric field and $B^i=-\epsilon_{ijk} F^{jk}/2$ the dark magnetic field; $\phi (t,\bm{r},\bm{R})$ is the bilocal field of the dark fermion-antifermion pair and its equation of motion is a Schrödinger equation with Hamiltonian
\begin{equation}
\begin{aligned}
 H(\bm{r},\bm{p},\bm{P},\bm{S}_1,\bm{S}_2) &= 2M + \frac{\bm{p}^2}{M}+\frac{\bm{P}^2}{4M} - \frac{\bm{p}^4}{4M^3} +  V (\bm{r},\bm{p},\bm{P},\bm{S}_1,\bm{S}_2) + \ldots \, ,
 \label{ham_pNRQED}
\end{aligned}
\end{equation}
\begin{equation}
\begin{aligned}
  &V (\bm{r},\bm{p},\bm{P},\bm{S}_1,\bm{S}_2)= V^{(0)} + \frac{V^{(1)}}{M} + \frac{V^{(2)}}{M^2} + \ldots \, ,
 \label{pot_pNRQED}    
\end{aligned}
\end{equation}
where $\bm{S}_1=\bm{\sigma}_1/2$ and $\bm{S}_2=\bm{\sigma}_2/2$ are the spin operators acting on the fermion and antifermion, respectively. At leading order the static potential is the Coulomb potential $V^{(0)}=-\alpha/r$; it scales like $M\alpha^2$, whereas the kinetic energy with respect to the relative motion, $\bm{p}^2/M$, scales as $M\alpha^2$ or $T$. For the center-of-mass kinetic energy it holds $\bm{P}^2/(4M) \sim T$ according to the hierarchy~\eqref{scale_arrang}, hence it is not suppressed. The DM pair can be decomposed in a discrete spectrum of bound states and a continuous spectrum of scattering states. At leading order in the center-of-mass frame, the energies are $E_n = 2M + E^b_n$ and $E_p = 2M + \bm{p}^2/M$, respectively, where $E^b_n=-M\alpha^2/(4n^2)$ is the binding energy.

Annihilation processes originate from the imaginary part of the contact potentials in pNRQED$_{\textrm{DM}}$ directly inherited from the four-fermion operators in NRQED$_{\textrm{DM}}$, cf.~\eqref{local_op_total_mom},
\begin{equation}
  \mathcal{L}_{\textrm{pNRQED}_{\textrm{DM}}} ~\supset~   -\int d^3r \; \phi^\dagger(t,\bm{r},\bm{R}) \; {\rm{Im}}\,\delta V^{\textrm{ann}} \; \phi (t,\bm{r},\bm{R})  \,,
\label{pNREFT_2}
\end{equation}
with 
\begin{equation}
\begin{aligned}
& {\rm{Im}} \,\delta V^{\textrm{ann}}
  = - \frac{1}{M^2} \, \delta^3(\bm{r})\, \left[ 2 {\rm{Im}}\,d_s - \bm{S}^2 \left( {\rm{Im}}\,d_s- {\rm{Im}}\,d_v \right) \right] \\
  & \; + \frac{1}{M^4} \, \delta^3(\bm{r})\, \nabla_{\bm{R}}^i \nabla_{\bm{R}}^j
  \left[ {2\rm{Im}}\, g_{\textrm{c\,cm}}\,\delta_{ij}
    - \bm{S}^2 \left( {\rm{Im}}\, g_{\textrm{c\,cm}} - {\rm{Im}}\,g_{\textrm{a\,cm}}\right)\delta_{ij}
    + S^i S^j \,{\rm{Im}}\,g_{\textrm{b\,cm}}  \right] \, .
\label{annpNRQED}
\end{aligned}
\end{equation}  
The operator $\bm{S}=\bm{S}_1+\bm{S}_2$ is the total spin of the dark fermion-antifermion pair.
\section{Annihilations and decays in the laboratory frame}
\label{sec:ann}
Concerning scattering states above the mass threshold, we compute the spin averaged annihilation cross section in the laboratory frame, using the optical theorem,
\begin{equation}
\sigma_{\hbox{\scriptsize ann}} v_{\hbox{\tiny M\o l}}  = \frac{{\rm{Im}}[\mathcal{M}_{\hbox{\tiny NR}}(\phi \to \phi)]}{2} \, ,
\label{optical_cross_section2}
\end{equation}
where the amplitude $\mathcal{M}_{\hbox{\tiny NR}}(\phi \to \phi)$ describes the propagation of the fermion-antifermion field $\phi$ projected on scattering states. For S-wave annihilation at leading order in $\alpha$ and at order $\bm{P}^2/M^2$ in the center-of-mass momentum, we obtain
\begin{equation}
  \left(\sigma_{\hbox{\scriptsize ann}} v_{\hbox{\tiny M\o l}}\right)_{\textrm{lab}}(\bm{p},\bm{P})
=  \sigma^{\hbox{\tiny NR}}_{\hbox{\scriptsize ann}} v^{(0)}_{\hbox{\scriptsize rel}}  \, S_{\hbox{\scriptsize ann}}(\zeta)\left(1- \frac{\bm{P}^2}{4M^2}\right) \, ,
\label{ann_fact_scat}
\end{equation}
where $S_{\hbox{\scriptsize ann}}(\zeta) \equiv  \left|\Psi_{\bm{p} 0}(\bm{0})\right|^2$
is the Sommerfeld factor~\cite{Sommerfeld} and $\sigma^{\hbox{\tiny NR}}_{\hbox{\scriptsize ann}} v^{(0)}_{\hbox{\tiny rel}} = ({{\rm{Im}}\,d_s}+3 {{\rm{Im}}\,d_v})/M^2=\pi \alpha^2/M^2$ the S-wave annihilation cross section for a free DM pair. As expected, the result in eq.~\eqref{ann_fact_scat} corresponds to a Lorentz-contracted cross section when going from the center-of-mass to the laboratory frame, i.e. $\left(\sigma_{\hbox{\scriptsize ann}} v_{\hbox{\tiny M\o l}}\right)_{\textrm{lab}}=\left(\sigma_{\hbox{\scriptsize ann}} v_{\hbox{\tiny M\o l}}\right)_{\textrm{cm}}/\gamma^2$, where $\gamma = 1/\sqrt{1-\bm{v}^2} \approx 1+\bm{P}^2/(8M^2)$ is the Lorentz boost factor. In fact, since the cross section $\sigma$ is in general Lorentz invariant in particle physics, one would obtain eq.~\eqref{ann_fact_scat} by simply transforming the M\o ller velocity up to first order in the center-of-mass momentum,
\begin{equation}
\begin{aligned}  
  \left(v_{\hbox{\tiny M\o l}}\right)_{\textrm{lab}} 
  = \frac{\left(v_{\hbox{\tiny M\o l}}\right)_{\textrm{cm}}}{\gamma^2} \approx \left(v_{\hbox{\tiny M\o l}}\right)_{\textrm{cm}}\left(1- \frac{\bm{P}^2}{4M^2}\right) .
  \label{MolLor}
\end{aligned}  
\end{equation}
Below the mass threshold, the decay width of a spin-singlet or spin-triplet bound state, called para- or orthodarkonium, can be computed from the optical theorem,
\begin{equation}
\Gamma_{\hbox{\scriptsize ann}} = 2 \,{\rm{Im}}[\mathcal{M}_{\hbox{\tiny NR}}(\phi \to \phi)] \,.
\label{optical_cross_section3}
\end{equation}
At leading order in the coupling only spin-singlet S-wave bound states decay into two dark photons. At order $\bm{P}^2/M^2$, the paradarkonium decay width in the laboratory frame reads
\begin{equation}
  \left(\Gamma^{n,\hbox{\scriptsize para}}_{\textrm{ann}}\right)_{\textrm{lab}}(\bm{P}) =
  \frac{4 {\rm{Im}}\,d_s}{M^2} \left|\Psi_{n00}(\bm{0})\right|^2 \left(1- \frac{\bm{P}^2}{8M^2}\right) \,,
\label{ann_para}
\end{equation}
where the square of the bound-state wavefunction $\left|\Psi_{n00}(\bm{0})\right|^2$ is given in the center-of-mass frame. Eq.~\eqref{ann_para} simply expresses the expected Lorentz dilation of time intervals up to first order in the non-relativistic expansion, $\left(\Gamma^{n,\hbox{\scriptsize para}}_{\textrm{ann}}\right)_{\textrm{lab}} = \left(\Gamma^{n,\hbox{\scriptsize para}}_{\textrm{ann}}\right)_{\textrm{cm}}/\gamma$.

\section{Formation of bound states in the laboratory frame}
\label{sec:bsf}
\begin{figure}[ht]
    \centering
    \includegraphics[scale=0.875]{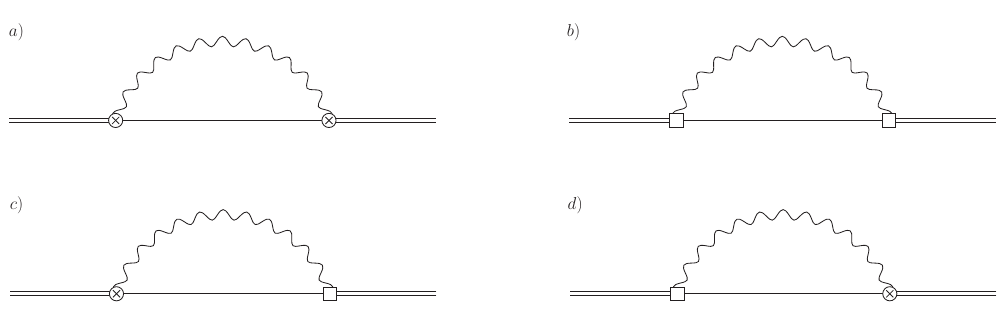}
    \caption{Self-energy diagrams in  pNRQED$_{\textrm{DM}}$ with an initial scattering state (solid double line) and an intermediate bound state (solid line) in the laboratory frame.
      Electric and magnetic couplings are represented by a circle-crossed and square vertex, respectively.}
    \label{fig:pnEFT_DM_self}
\end{figure}

In the laboratory frame, the center of mass of the dark matter pair is moving, whereas the thermal bath is at rest. The formation of a bound state out of a scattering state, which emits a dark photon carrying energy and momentum of order $M\alpha^2$ or $T$, can be computed at order $\bm{r}^2$ from the two interaction terms in eq.~\eqref{pNREFT_1}, by cutting at finite $T$ the time-ordered self-energy diagrams displayed in figure~\ref{fig:pnEFT_DM_self}. 
They depend on the electric-electric correlator $\langle E_i(t,\bm{R})E_j(0,\bm{R}') \rangle$ (diagram a)), the magnetic-magnetic correlator $\langle B_i(t,\bm{R})B_j(0,\bm{R}') \rangle$ (diagram b)), the electric-magnetic correlator $\langle B_i(t,\bm{R})E_j(0,\bm{R}') \rangle$ (diagram c)) and the magnetic-electric correlator $\langle E_i(t,\bm{R})B_j(0,\bm{R}') \rangle$ (diagram d)). Since the dark photon in the loop carries a spatial momentum $\bm{k}$, the bound fermion-antifermion pair recoils by a spatial momentum $\bm{P}-\bm{k}$, resulting in a recoil correction $(2\bm{P}\cdot \bm{k}-\bm{k}^2)/(4M)$ in the bound-state propagator. It is suppressed by at least $\sqrt{T/M}$ with respect to $\Delta E^{p}_n\equiv \bm{p}^2/M-E^b_n$ and $T$, hence we can expand the propagator in the recoil term.

The sum of the imaginary parts of the self-energies a)-d) gives the bound-state formation cross section up to relative order $\bm{P}^2/M^2 \sim T/M$ and $\Delta E^p_n/M$ in the laboratory frame:
\begin{equation}
\begin{aligned}
  &(\sigma_{\hbox{\scriptsize bsf}} \, v_{\hbox{\scriptsize M\o l}})_{\textrm{lab}}(\bm{p},\bm{P}) \equiv  \sum \limits_n   (\sigma^n_{\hbox{\scriptsize bsf}} \, v_{\hbox{\scriptsize M\o l}})_{\textrm{lab}}(\bm{p},\bm{P})
  = -2\sum_{i=a,b,c,d}\textrm{Im} \left(\Sigma^{11}_{\textrm{diag.}\,\textit{i)}}\right)_{\textrm{lab}}(\bm{p},\bm{P}) \\
  & \hspace{0.5cm} =\frac{4}{3}\alpha \sum \limits_n  (\Delta E^p_n)^3 \left(1+n_{\textrm{B}}( \Delta E^p_n)\right)
 \left( |\langle n|\bm{r}|\bm{p}\rangle|^2 F^n_1(p,P) + \left|\langle n|\bm{r}\cdot \frac{\bm{P}}{2M}|\bm{p}\rangle\right|^2 F^n_2(p,P)\right) \, ,
\label{bsf_xsection}
\end{aligned}
\end{equation}
where
\begin{equation}
\begin{aligned}
  F^n_1(p,P) =&\,  1 - \frac{3}{4}\frac{\Delta E^p_n }{M}+\frac{\bm{P}^2}{4M^2} + n_{\textrm{B}}( \Delta E^p_n) \,\frac{\Delta E^p_n}{4M} \, \frac{\Delta E^p_n}{T}\\
  &\, - n_{\textrm{B}}( \Delta E^p_n) \, \frac{\bm{P}^2}{4M^2}
     \frac{\Delta E^p_n}{T} \left[ 1 - \frac{\Delta  E^p_n}{5T}  - \frac{2}{5} n_{\textrm{B}}( \Delta E^p_n) \frac{\Delta E^p_n}{T} \right],
  \label{bsf_xsection_final_result_F1}
\end{aligned}
\end{equation}
and
\begin{equation}
  F^n_2(p,P) =  1 - \frac{1}{10} \, n_{\textrm{B}}( \Delta E^p_n) \, \frac{(\Delta E^p_n)^2}{T^2} \, \left( 1+ 2 \, n_{\textrm{B}}( \Delta E^p_n) \right).
\label{bsf_xsection_final_result_F2}
\end{equation}
Note that the statistical factor $1+n_{\textrm{B}}( \Delta E^p_n)$ in \eqref{bsf_xsection}, where $n_{\textrm{B}}(E)=1/(e^{E/T}-1)$ is the Bose--Einstein distribution function, reflects the fact the the dark photon is emitted into the thermal bath. In ref.~\cite{biondini2024effective}, the relation between the bound-state formation cross section in the laboratory frame, cf.~\eqref{bsf_xsection}, and the expression in the center-of-mass frame, where the dark pair is at rest and the thermal medium moves with non-relativistic velocity $\bm{v}=-\bm{P}/(2M)$, is shown to fulfill the same Lorentz transformation formula as in the case of annihilations,
\begin{equation}
  (\sigma_{\hbox{\scriptsize bsf}} \, v_{\hbox{\scriptsize M\o l}})_{\textrm{lab}}(\bm{p},\bm{P}) = 
  (\sigma_{\hbox{\scriptsize bsf}} \, v_{\hbox{\scriptsize M\o l}})_{\textrm{cm}}(\bm{p},\bm{v}) \left(1 - \frac{\bm{P}^2}{4M^2} \right)\,.
\label{Lorentz_boost2}
\end{equation}
\section{Conclusions and outlook}
\label{sec:outlook}
In this proceeding, we have summarized the findings of the recent work~\cite{biondini2024effective}, where we use the language of NRQED and pNRQED to determine the annihilation cross section, decay width and bound-state formation cross section in the laboratory frame, relevant for the evolution of heavy thermalized DM pairs at and after the chemical freeze-out. In eqs.~\eqref{ann_fact_scat} and~\eqref{ann_para} we recover in the non-relativistic expansion the expected Lorentz dilation formulas for the annihilation cross section and decay width, respectively, due to the non-relativistic motion of the center of mass of the heavy pairs. Under the hierarchy of energy scales~\eqref{scale_arrang}, we compute the recoil corrections to the bound-state formation cross section up to order $\bm{P}^2/M^2 \sim T/M$ and $\Delta E^p_n/M$ in eq.~\eqref{bsf_xsection}. 

We observe that for times where the hierarchy~\eqref{scale_arrang} is certainly fulfilled, corrections due to the center-of-mass motion are within a few percentage and hence quite suppressed. For increasing temperatures, such that $T\gtrsim M\alpha^2$, the recoil corrections have a bigger impact on the annihilation cross section, decay width and the bound-state formation cross section. They are, however, suppressed when considering the dark matter evolution for most of the times after the chemical freeze-out. While for the model~\eqref{lag_mod_0} considered in this work we may extend the hierarchy up to temperatures $T \lesssim M\alpha$, where eq.~\eqref{bsf_xsection} still gives the dominant contribution to the formation of bound states, in non-abelian dark gauge field theories the hard thermal loop resummation of the Debye mass scale must be taken into account. It is left for a future work to study the center-of-mass motion in such theories with an additional energy scale.

\bibliographystyle{JHEP.bst}
\bibliography{PoS_proceeding.bib}

\end{document}